\documentclass[fleqn,usenatbib]{mnras}

\usepackage{graphicx}
\usepackage{longtable}
\usepackage{float}
\usepackage{amssymb,amsmath}
\usepackage{listings}
\usepackage{xcolor}
\usepackage{CJK}
\usepackage{threeparttable}
\usepackage{lastpage}
\usepackage{prettyref}

\hypersetup{
    colorlinks,
    linkcolor={red!50!black},
    citecolor={blue!50!black},
    urlcolor={blue!80!black}
}

\newcommand{\dd}{{\rm d}}
\newcommand{\gcmc}{$\mathrm{g\,cm^{-3}}$}
\newcommand{\magra}{MAGRATHEA}
\newcommand{\gru}{Gr\"{u}neisen}
\graphicspath{{../figures/}}

\title[\magra]{\magra: an open-source spherical symmetric planet interior structure code}
\author[C. Huang et al.]{
Chenliang Huang (黄辰亮),$^{1}$\thanks{E-mail: huangcl@arizona.edu, david.rice@unlv.edu}
David R. Rice $^{2}$\footnotemark[1]
and Jason H. Steffen$^{2}$
\\
$^{1}$Lunar and Planetary Laboratory, University of Arizona, Tucson, AZ 85721\\
$^{2}$Department of Physics \& Astronomy, University of Nevada Las Vegas, PO Box 454002, Las Vegas, NV 89154
}

\begin{document}
\begin{CJK*}{UTF8}{gkai}

\label{firstpage}
\pagerange{\pageref{firstpage}--\pageref{lastpage}}
\maketitle

\begin{abstract}

\magra\ is an open-source planet structure code that considers the case of fully differentiated spherically symmetric interiors.  Given the mass of each layer and the surface temperature, the code iterates the boundary conditions of the hydrostatic equations using the method of shooting to a fitting point in order to find the planet radius.  The first version of \magra\ supports a maximum of four layers of iron, silicates, water, and ideal gas.  With a few exceptions, the temperature profile can be chosen between isothermal, isentropic, and user-defined functions.  The user has many options for the phase diagram and equation of state in each layer and we document how to add additional equations of state.  We present \magra's capabilities and discuss its applications.  We encourage the community to participate in the development of \magra\ at \url{https://github.com/Huang-CL/Magrathea}. 
\end{abstract}

\begin{keywords}
planets and satellites: composition -- planets and satellites: general -- planets and satellites: interiors -- equation of state
\end{keywords}

\section{Introduction}
\label{sec:intro}
Discoveries such as Kepler-10b, LHS 1140b, the seven planets of the TRAPPIST-1 system, and Proxima Centauri b \citep{kepler10b,lhs1140b,LHS1140,trappist1first,TRAPPIST1,proximab} show that the discovery of small rocky planets, or planets with minimal atmospheres, are now firmly within observational capabilities.  Data from the Kepler space telescope shows that planets between 1-4 R$_\oplus$ are among the most common types in our galaxy \citep{Batalha2013, Fressin2013, Petigura2013}.  The densities of these observed terrestrial planets allow for a range of interior compositions from Mercury-like \citep{K2-229} to volatile rich \citep{Leger04}.  Fitting mass-radius relationships to observed planets provides a probe into planet composition (e.g. \citet{Zeng2013,Unterborn2018}).  However, the amount of mass in each differentiated layer cannot be determined by density alone \citep{Rogers2010, Dorn2015}.

The community uses a variety of models to characterize the interior structure of planets.  Underlying these models are differing computational techniques (e.g. shooting in \citet{Nixon2021}, relaxation in \citet{Unterborn2018}, or calculus of variation in \citet{Zeng2021}) and numerous experimental measurements and theoretical estimates of the equation of state (EOS) for planet-building materials.  For example, a number of models use bridgmanite primarily in the mantle and do not include high-pressure post-perovskite \citep{Grasset2009, Zeng2016, Brugger2017} while others capture detailed upper-mantle chemistry \citep{Valencia2007, Dorn2017, Unterborn2019} and recent work adds liquid silicates \citep{Noack2020}.  Another point of difference is the model's temperature profile with some using an isothermal mantle \citep{Seager2007} rather than isentropic \citep{Hakim2018}.  While others limit isothermal modeling to the water layer \citep{Sotin2007, Brugger2016} and the atmosphere \citep{Madhusudhan2021, Baumeister2020}.  Newer models try to capture the complex phase diagram of water \citep{Mousis2020, Journaux2020, Haldemann2020}, the hydration of mantle materials \citep{Shah2021, Dorn2021}, and couple the atmosphere to the interior \citep{Madhusudhan2020, Acuna2021}.

We develop \magra\footnote{Magrathea is the legendary planet where hyperspatial engineers manufacture custom-made planets in Douglas Adams's The Hitchhiker's Guide to the Galaxy \citep{hitchikers}.} an open-source planet interior solver that can be customized to different user-defined planet models.  Compared to other codes, the package is designed to enhance ease-of-use and flexibility.  \magra\ features phase diagram options and transparent EOS formatting, which enables the user to choose between a large library of EOSs and add/change materials and equations. 

We are motivated by our collaboration with high-pressure physicists to understand how uncertainties in experimental EOSs affect predictions of planet interiors.  In \citet{Huang21}, we use our adaptable planet interior model to implement new measurements for high-pressure water-ice from \citet{Grande2022}.  Their experiments confirmed the pressure of the transition from ice-VII to ice-X and identified a transitional tetragonal ice-VII$_\textrm{t}$ phase.  This improved H$_2$O equation of state changed the predicted radius of a pure water, 10 M$_\oplus$ planet by over four per cent from \citet{Zeng2016}.

In this paper, we document the features and demonstrate the functionality of \magra. The most up-to-date version is hosted on the GitHub platform.  Our paper is laid out as follows.  We describe the fundamentals of a planet interior solver in Section \ref{sec:model}.  In Section \ref{sec:overv-code-struct}, we describe the specifics of our code and how to build a planet model within the code.  A model is designated by defining a phase diagram for each layer and choosing an equations of state from our library for each phase.  We discuss limitations to our default model in \ref{sec:limitations}.  We describe the code's functionality and discuss various tests in Section \ref{sec:test-problems}, and end with some summarizing remarks in Section \ref{sec:summary}.

\section{Interior Structure Solver}
\label{sec:model}

\begin{figure*}
    \centering
    \includegraphics[width=\textwidth]{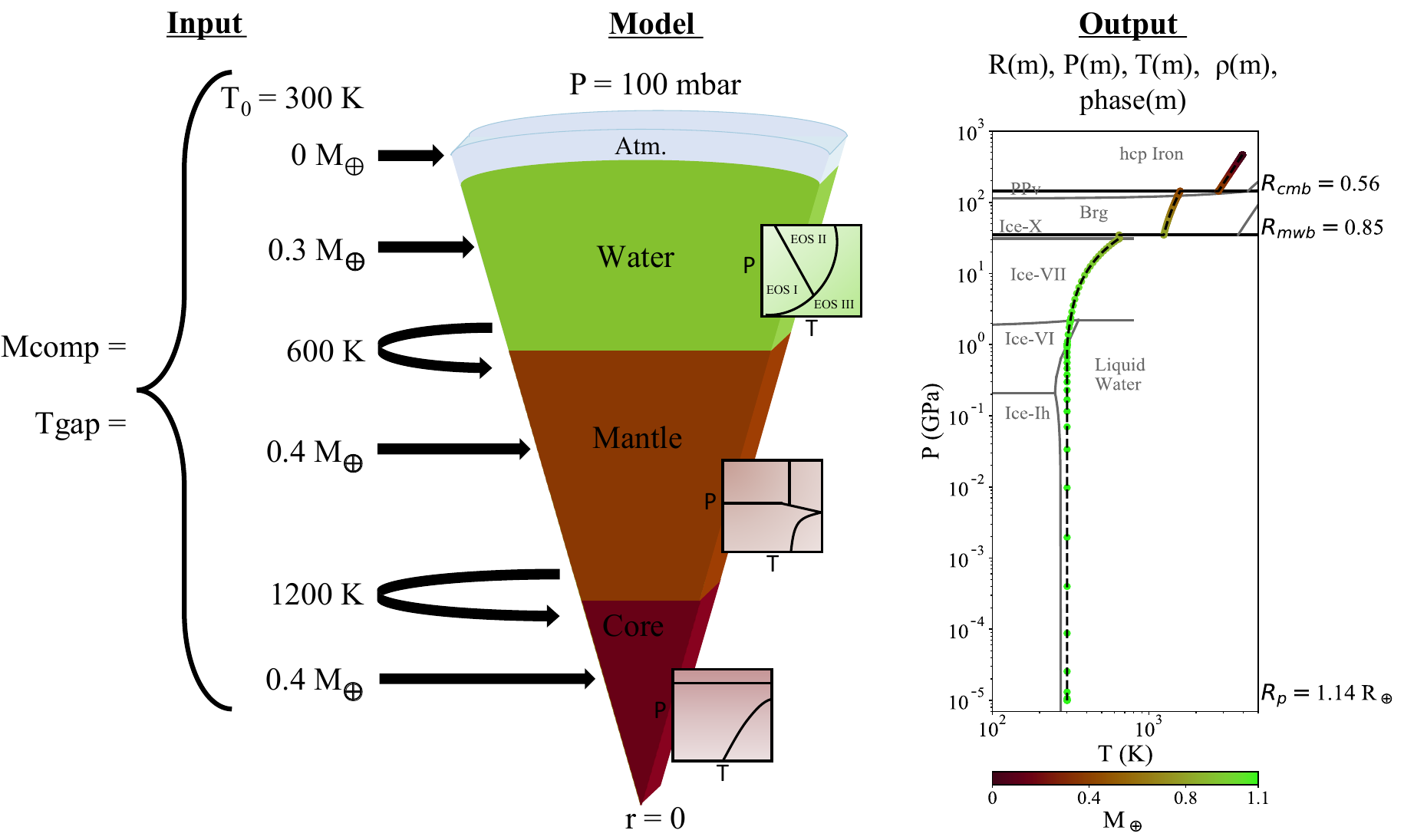}
    \caption{A schematic overview of \magra. Showing an example input, \textit{left}, of a 1.1 M$_\oplus$ planet with 0.4 M$_\oplus$ core, 0.4 M$_\oplus$ mantle, and 0.3 M$_\oplus$ hydrosphere.  The planet is not in thermal equilibrium with a surface temperature of 300 K and jumps in temperature across boundary layers of 600 K and 1200 K.  \textit{Center}, shows \magra's four input layers with cartoons of phase diagrams defined for each layer with an EOS chosen for each phase.  Default phase diagrams shown in Fig.~\ref{fig:phasediagram}.  \textit{Right}, shows the pressure and temperature with enclosed mass.  The radius at boundaries and the planet radii is also shown.}
    \label{fig:schematic}
\end{figure*}

We consider a simplified planet structure with four layers: an iron core, a silicate-dominated mantle, a hydrosphere of water/ice, and an ideal gas atmosphere.  We assume these layers are spherically symmetric and that a single solution to composition at a given pressure and temperature exists within each layer.  Our approach is similar to other interior structure solvers used for exoplanets (e.g.~\citealt{Leger04,Valencia2006,Sotin2007,Seager2007}).  A schematic depiction of \magra\ from input to output is shown in Fig.~\ref{fig:schematic}.  Given the mass of each of these layers, $M_{\text{comp}}=\{M_{\text{core}},M_{\text{mantle}},M_{\text{hydro}},M_{\text{atm}}\}$, the code calculates the radius returning the pressure $P(m)$, density $\rho(m)$, and temperature $T(m)$ with enclosed mass $m$ by solving the following four equations:

\begin{enumerate}
\item Mass continuity equation
\begin{equation}
  \label{eq:1}
  \frac{\dd r(m)}{\dd m} = \frac{1}{4 \pi r^2 \rho(m)},
\end{equation}

\item Hydrostatic equilibrium
\begin{equation}
  \label{eq:2}
  \frac{\dd P(m)}{\dd m} = -\frac{G m}{4 \pi r^4},
\end{equation}

\item Temperature gradient\\
If the isothermal option is chosen, the temperature gradient is 0.  When the isentropic option is chosen, depending on the available thermal properties of the phases, the temperature gradient can be calculated using either of the following two formulae. 
If the \gru\ parameter $\gamma$ is available, we can have
\begin{align*}
  \frac{\dd T(m)}{\dd m} &= \frac{\dd T}{\dd V}\Bigr|_S \frac{\dd V}{\dd \rho}\frac{\dd \rho}{\dd m} \\
  &=-\frac{m_{\rm mol}}{\rho^2}\frac{\dd T}{\dd V}\Bigr|_S\left(\frac{\partial \rho}{\partial P}\frac{\dd P}{\dd m}+\frac{\partial \rho}{\partial T}\frac{\dd T}{\dd m}\right).
\end{align*}
 Thus,
 \begin{equation}
 \label{eq:dTdm}
     \frac{\dd T(m)}{\dd m} = \frac{\frac{\dd T}{\dd V}\Bigr|_S\frac{Gm}{4\pi r^4}}{\frac{\rho^2}{m_{\rm mol}}\frac{\partial P}{\partial \rho}\Bigr|_T -\frac{\dd T}{\dd V}\Bigr|_S\frac{\partial P}{\partial T}\Bigr|_\rho},
 \end{equation}
 where
 \begin{equation}
 \label{eq:last}
    \frac{\dd T}{\dd V}\Bigr|_S = - \frac{\gamma T}{V},
 \end{equation}
$m_{\rm mol}$ and $V$ are the molar mass and volume respectively.

Alternatively, if the thermal expansion $\alpha$ is available, we can have
\begin{equation}
\label{eq:dTdm_alpha}
    \frac{\dd T(m)}{\dd m} = -\frac{\alpha T G m}{4 \pi r^4 \rho c_p},
\end{equation}
where $c_p$ is the specific heat capacity at constant pressure.

\item Equation of state (EOS)
  \begin{equation}
  \label{eq:eos}
  P(m) = P(\rho(m),T(m)),
\end{equation}
which is unique for each material/phase.
\end{enumerate}

The boundary conditions of the model are $r=0$ at $m=0$ and the user-defined surface temperature and pressure, $T(M_{p})$ and $P(M_{p})$.  The default surface pressure is 100 mbar approximately the pressure level of the broad-band optical transit radius probes \citep{TRAPPIST1}.  Since a rapid temperature jump may occur at the boundary between chemically distinct layers, where heat can only transfer conductively \citep{Valencia2006}, the user can also set a temperature discontinuity, $T_{\text{gap}}$, at the boundary of adjacent layers.  The Earth is often modeled with a discontinuity ranging from a 300 to 1100 K increase across the core mantle boundary \citep{Nomura2014,Lay2008} and \citep{Valencia2007} and \citep{Sotin2007} use discontinuities in their exoplanet models.  In our demonstrations in Section \ref{sec:test-problems}, we use planets that are in thermal equilibrium between their layers with $T_{\text{gap}}=\{0,0,0,T(M_{p})\}$.

The user can choose between isothermal and isentropic temperature mode by adjusting an {\it isothermal} flag of the integrator.  When the flag is set to true, the whole planet is assumed to be isothermal (except the atmosphere layer, see section~\ref{sec:gas}).  When the input EOS parameters of a phase are not sufficient to calculate the isentropic temperature gradient (see section~\ref{sec:formula}), this specific phase would be treated as isothermal, regardless of the value of the {\it isothermal} flag.

\subsection{Phase diagrams}
\label{sec:Phaseoverview}

We model our planets with distinct differentiated layers with defined mass fractions.  However, within each layer the phase may change due to the large pressure and temperature ranges.  When integrating within a layer, the code first checks the pressure and temperature to determine the appropriate region of the phase diagram.  A built-in phase function outputs a link to the EOS of the material or phase in the region.  The density $\rho$ or the volume of unit cell $V$ can then be solved from the EOS (Eq.~\ref{eq:eos}).

\begin{figure*}
    \centering
    \includegraphics[width=0.3\textwidth]{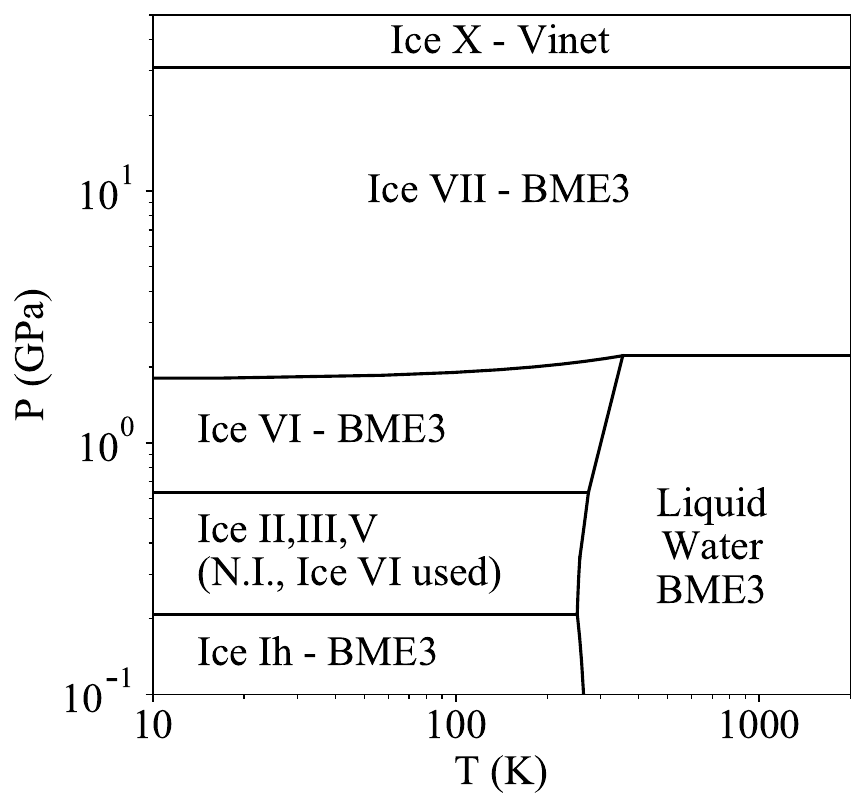}
    \includegraphics[width=0.3\textwidth]{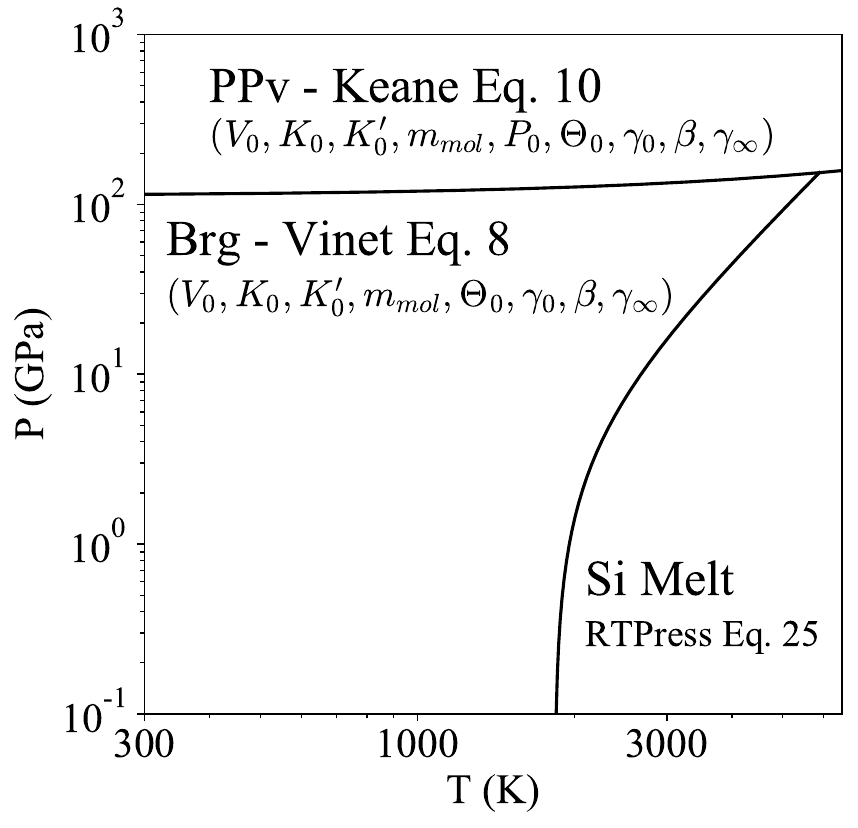}
    \includegraphics[width=0.3\textwidth]{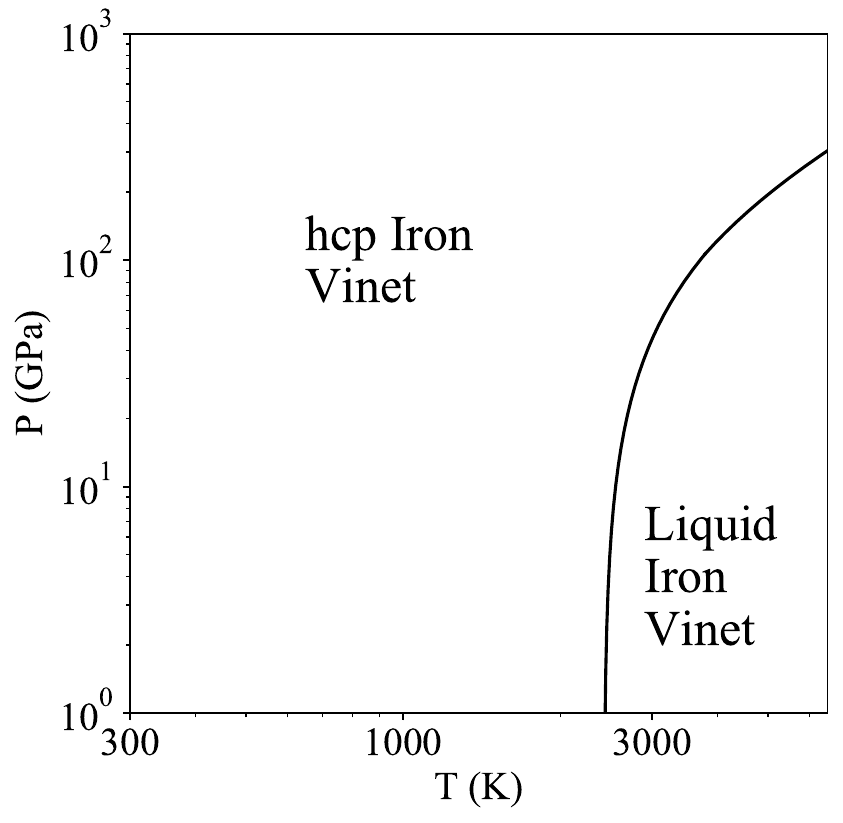}
    \caption{Default phase diagrams for hydrosphere, mantle, and core layers.  EOS and phase transitions from a variety of sources detailed in Sec.~\ref{sec:eoslist}.  Type of EOS fitting equation shown from Sec.~\ref{sec:formula}.  Additionally, the parameters for mantle EOSs are shown, \textit{middle}.  The phase diagrams and choice of EOS can be customized by the user.  \textit{N.I.} is \textit{not implemented}.  The atmosphere layer also has a phase function, but our default is ideal gas at all pressure and temperatures, see Sec.~\ref{sec:gas}}
    \label{fig:phasediagram}
\end{figure*}

A key feature of \magra\ is the user's ability to change the phase diagram in each layer and choose between EOSs for each phase.  Fig.~\ref{fig:phasediagram} shows our default phase diagram and Sec.~\ref{sec:phase} further details their implementation.  The atmosphere layer is not shown in the plots and has two ``phases'', isothermal for P$<$100 bar and adiabatic for P$>$100 bar.

\subsection{Equation of state formulae}
\label{sec:formula}

The density of a phase at certain temperature and pressure is determined by the EOS, which can be fed into \magra\ either through an analytical fitting formula or a tabulated pressure-density table.  For the first option, each material's EOS must provide the parameters for one of the following formulations. 

According to the Mie-Gr\"{u}neisen-Debye (MGD) formulation, the total pressure $P$ can be divided into an isothermal term $P_c$, and a thermal term, $P_{th}$, expressed as \citep{Dewaele06}:
\begin{equation}
  \label{eq:4}
  P(V,T) = P_c(V) + P_{th}(V,T) - P_{th}(V,T_0)
\end{equation}
\magra\ includes four common types of the EOS that give the reference isotherm $P_c$, including:

\begin{enumerate}
\item The Eulerian finite strain Birch-Murnaghan EOS (BME) is the most commonly used EOS. The fourth-order BME \citep{Seager2007} is  
\begin{align}
  \label{eq:bme}
  P_c &= \frac{3}{2}K_0 \left(\eta^{7/3}-\eta^{5/3}\right) \\ \nonumber
  &\left(1+\zeta_1(1-\eta^{2/3}) +\zeta_2(1-\eta^{2/3})^2 \right),
\end{align}
where
\begin{equation*}
  \eta = V_0/V, \zeta_1 = \frac{3}{4}(4-K_0')
\end{equation*}
\begin{equation*}
  \zeta_2 = \frac{3}{8}\left(K_0K_0''+K_0'(K_0'-7)+\frac{143}{9}\right).
\end{equation*}

$K=-V(\partial P/\partial V)_T$ is the isothermal bulk modulus, $K'$ is the first derivative of the bulk modulus with respect to pressure, and $K''$ is the second of the bulk modulus with respect to pressure.  The subscript 0 refers to quantities at ambient-pressure conditions.  If $\zeta_2$ is set to zero the EOS reduces to third-order BME.

\item Vinet EOS \citep{Seager2007,Smith2018,Wicks2018}, which is considered to give more accurate result than BME at high pressure or large compression \citep{Poirier00}
\begin{align}
  \label{eq:vinet}
  P_c &= 3K_0 \eta^{2/3}\left(1-\eta^{-1/3}\right) \\ \nonumber
 &\exp\left(\frac{3}{2}(K_0'-1)\left(1-\eta^{-1/3}\right)\right);
\end{align}

\item Holzapfel EOS \citep{Bouchet13}
\begin{align}
  \label{eq:holz}
  x &= (V/V_0)^{1/3} \\ \nonumber
  c_0 & = - \ln \left(\frac{3K_0}{1003.6{\rm~GPa~cm^5~mol^{-5/3}}(Z/V_0)^{5/3}}\right) \\ \nonumber
  c_2 & = \frac{3}{2}(K_0'-3)-c_0 \\ \nonumber
  P_c & = 3K_0 x^{-5}(1-x) \\ \nonumber
 &\exp\left(c_0(1-x)\right)\left(1+c_2x(1-x)\right);
\end{align}

\item Keane EOS \citep{Sakai16}
\begin{align}
  \label{eq:keane}
  y & = V_0/V \\ \nonumber
  K_\infty' &= 2\left(\gamma_\infty+\frac{1}{6}\right) \\ \nonumber
  P_c & = \frac{K_0' K_0}{K_\infty'^2}\left(y^{K_\infty'}-1\right)-(K_0'-K_\infty')\frac{K_0}{K_\infty'}\ln(y)
\end{align}

\item An empty placeholder is available for an additional formulation of the EOS such as from \citet{Choukroun2007}.
\end{enumerate}

For phases that only exist at high pressure, the bulk modulus, $K_P$, can be measured more precisely at the phase transition pressure, $P_0$, than at ambient pressure~\citep{Salamat13}.  If $K_P$ is applied instead of $K_0$, a constant pressure $P_0$ under which the bulk modulus is measured is added to the isotherm pressure.  If the parameters to calculate the thermal term, $P_{th}$, are not provided, only $P_c$ is calculated and a constant temperature is returned.

$P_{th}$ is most commonly calculated from a quasi-harmonic Debye thermal pressure and the anharmonic and electronic thermal pressure, which are obtained from \citep{Belonoshko08,Duffy15}
\begin{equation}
  \label{eq:9}
  P_{th}(T)=\frac{\gamma}{V}E_{th}(T)+\frac{3nR}{2V}e_0x^ggT^2,
\end{equation}
where 
\begin{equation}
  \label{eq:12}
  x = V/V_0 = \eta^3, \gamma = \gamma_\infty+(\gamma_0-\gamma_\infty)x^\beta
\end{equation}
\begin{equation}
  \label{eq:13}
  \Theta = \Theta_0 x^{-\gamma_\infty}\exp\left(\frac{\gamma_0-\gamma_\infty}{\beta}(1-x^\beta)\right)
\end{equation}
\begin{equation}
  \label{eq:14}
  z = \Theta/T, E_{th}=3nRTD_3(z). 
\end{equation}
$D_3$ indicates a third-order Debye function. $\Theta_0$, $\gamma_\infty$, $\gamma_0$, $\beta$, $e_0$, and $g$ are fitting parameters, $R$ is the gas constant, and $n$ is the number of atoms in the chemical formula of the compound.  A numerical derivative of Eq.~\ref{eq:4} gives the $\frac{\partial P}{\partial \rho}\Bigr|_T$ and $\frac{\partial P}{\partial T}\Bigr|_\rho$ that are required in the Eq.~\ref{eq:dTdm}.

A different framework for the thermal term, referred to as RTpress in the following, is used in \citet{Wolf2018} and built upon the Rosenfeld-Trazona model.  In addition to the \gru\ parameters, it has additional fitting terms with derivation in \citet{Wolf2018}.
 
In RTpress, the \gru\ parameter can be written as 
\begin{equation}
    \gamma = \gamma_{\rm 0S}\frac{C_{\rm V}(V,T_{0S})}{C_V(V,T)}+V\frac{b'(V)}{b(V)}\frac{\Delta S^{\rm pot}(T_{0S}\to T)}{C_V(V,T)}.
\end{equation} 
  The subscript 0S stands for the quantity evaluated along the reference adiabat, 
  \begin{equation}
      T=T_{0S}=T_0\sqrt{1+a_1f+\frac{1}{2}a_2f^2},
  \end{equation}
 with,
 \begin{equation}
     f\equiv f(V)=\frac{1}{2}\left[\left(\frac{V_0}{V}\right)^{\frac{2}{3}}-1\right], 
 \end{equation}
 
 \begin{equation} \label{eq:RTpress_gru}
    a_1=6\gamma_0, \quad {\rm and } \quad a_2=-12\gamma_0+36\gamma_0^2-18\gamma_0',
 \end{equation}
 where $\gamma_0$ is the \gru\ parameter at zero GPa and $\gamma_0'=V_0(d\gamma/dV)_0$.
 
 The \gru\ parameter variation along the reference adiabat is   
 \begin{equation}
    \gamma_{0S}=\frac{(2f+1)(a_1+a_2f)}{6(1+a_1f+\frac{1}{2}a_2f^2}. 
 \end{equation}

\begin{equation}
\label{eq:bV}
b(V)=\sum_{n} b_{n} \left(\frac{V}{V_{0}}-1\right)^{n},
\end{equation}
and
\begin{equation}
b^\prime(V)=\sum_{n} b_{n} \left(\frac{n}{V_{0}}\right) \left(\frac{V}{V_{0}}-1\right)^{n-1}
\end{equation}
are a polynomial representation of the thermal coefficients and its volume derivative in cgs units. 

The total heat capacity 
 \begin{equation}
C_{V}(V, T) =b(V) f_{T}^{(1)}+\frac{3}{2} n R 
 \end{equation}
 is the sum of potential and kinetic contributions\footnote{We belive that Eq.~B.3 in \cite{Wolf2018} should read $C_V^{\rm kin}=\frac{3}{2}k_B$. To be consistent with the unit eV/atom chosen for $b_n$ in their work, the number of atoms per formula unit should not be a factor.}
 , where
 \begin{equation}
 \label{eq:fT}
 f_{T}=\left(\frac{T}{T_{0}}\right)^{\beta}-1, \quad f_{T}^{(1)}=\frac{\beta}{T_{0}}\left(\frac{T}{T_{0}}\right)^{\beta-1}.
\end{equation}

The difference in entropy from the reference adiabat from the potential contribution is
\begin{equation}
\Delta S^{\rm pot}\left(T_{0 S} \rightarrow T\right)=\frac{b(V)}{\beta-1} \left(f_{T}^{(1)}(T)-f_{T}^{(1)}\left(T_{0 S}\right)\right).
\end{equation}

Finally, the thermal term of the pressure in the framework of \citet{Wolf2018} can be written as 
\begin{align}
P_{th}=&-b^{\prime}(V) f_{T}+\gamma_{0 S}(V) \frac{C_{V, 0 S}(V)\cdot\left(T-T_{0}\right)}{V}  \nonumber \\ 
&+ \frac{b^{\prime}(V)}{\beta-1}\left[T \left( f_{T}^{(1)} - f_{T}^{(1)}(T_{0S})\right) \right. \nonumber \\
& \left. - T_0\left(f_{T}^{(1)}(T_{0}) - f_{T}^{(1)}(T_{0S})\right)\right]
\end{align}

Besides the MGD representation, the P-V-T equation of state can also be expressed as \citep{Fei1993}
\begin{equation}
    \label{eq:PVTalpha}
    V(P,T) = V(P,T_0)\exp\left[ \int^T_{T_0}\alpha(P,T')\dd T'\right],
\end{equation}
$\alpha(P,T)$ is the thermal expansion, which has the form of
\begin{equation}
\label{eq:alpha}
    \alpha(P,T)=\alpha(T)\left(1+\frac{K_0'}{K_0}P\right)^{-\xi}.
\end{equation}
$\alpha(T)$ is the zero-pressure thermal expansion coefficient. For $T>T_0$, it is expressed as a linear function of temperature
\begin{equation}
    \label{eq:alpha0}
    \alpha(T)=\alpha_0+\alpha_1T.
\end{equation}

For phases that adopt the thermal expansion representation, the Eq.~\ref{eq:dTdm_alpha} can be used to calculate adiabatic temperature gradient. The heat capacity in the equation can be calculated using the fitting formula 
\begin{equation}
    \label{eq:cp}
    c_p=c_{p0}+c_{p1}T-c_{p2}T^{-2}.
\end{equation}

\subsubsection{Tabulated Equation of State}
\label{sec:tabulated}

In place of a fitting equation, the EOS can take the form of a tabulated density-pressure table which then code then interpolates. For \magra\, the input file should have two columns, the first column is the density in g~cm$^{-3}$ and the second column is the pressure in GPa.  The pressure must be strictly ordered.  The first row of the table file, which contains header information, is skipped when the file is parsed.  The program interpolates the table monotonically using the Steffen spline \citep[no relation to our co-author]{steffenspline} from the $gsl$ package \citep{GSL}. 

\subsubsection{User-defined function}
\label{sec:user-defin-funct}

For further flexibility, users can create or modify the EOS of a phase using their own C++ functions.  \magra\ supports three types of user-defined functions.  First, a user can provide an EOS function that returns the material density as a function of pressure and temperature.  In addition, either a temperature gradient, $\dd T/\dd P$, as a function of pressure and temperature or an entropy function dependent on density and temperature can be used to set up the temperature solver in a phase.  If a user-defined temperature gradient function is set up, the temperature profile of this phase is not restricted to isentropic.  The density function and the temperature function ($\dd T/\dd P$ or entropy) can also be used in combination for a phase.  The method to import these user-defined functions is shown in Appendix~\ref{app:modify-eos}.

\section{Overview of the code structure}
\label{sec:overv-code-struct}

\magra\ is written in C++ and relies on the \textit{GNU Scientific Library (GSL)}\footnote{\textit{GNU Scientific Library} can be found at \url{http://www.gnu.org/software/gsl/}} \citep{GSL}.  A step by step guide to run the code can be found in the $README$ file on our GitHub repository.

The code is compiled with the included $Makefile$.  The central interaction with the user occurs through $main.cpp$.  The user may choose between seven modes by setting {\it input\_mode}.  They include the regular solver, a temperature-free solver, a two-layer solver (Sec.~\ref{sec:2layer}), three methods to change the EOS during run-time (Sec.~\ref{sec:eosunc}), and a bulk input mode for solving many planets with the regular solver in a single run (Sec.~\ref{sec:ternary}).  Each mode requires the user to define the mass of each layer in the planet.  In modes where the regular solver is used, the user defines a temperature array which gives the temperature at the surface and any discontinuities between layers.  This section covers the specific design of the code and options that the user may choose between---the solver in Sect.~\ref{sec:solver}, the phase diagram in ~\ref{sec:phase}, and the EOSs in Sec.~\ref{sec:eoslist}.

\subsection{Solver}
\label{sec:solver}

Solving the ordinary differential equations (ODE), Eq.~\ref{eq:1}-\ref{eq:last}, is a two-point boundary value problem \citep{Press2007}.  We have a total of three differential equations for $r$, $P$, and $T$ verses the independent variable $m$.  $P$ and $T$ need to satisfy boundary condition at $m=M_p$, and the boundary condition for $r$ is located at $m=0$.  The density may be discontinuous at a phase boundary whose location is unknown.  Therefore, in our case, \magra\ solves this problem with the method of shooting to a fitting point which is preferred to a relaxation method because it does not require predefined grids.

\magra\ integrates ODEs using GSL's $gsl\_odeiv2$ \citep{GSL} with a Runge-Kutta-Fehlberg method, which is a fourth order integrator with a fifth order error estimator and an automated adaptive step-size.  To have a better estimation on the boundary conditions, $P(m=0)$, $T(m=0)$, and $R_p$, we conduct an extra round of iteration using a ``pure'' shooting method.  In this first iteration, the ODEs are integrated from an estimated planet radius $R_p$ at $M_p$ outside-in toward the center until $m=0$ or $P(m)>10^{5}~\mathrm{GPa}$.  The $R_p$, which is the only unset boundary condition at $M_p$, is iterated using Brent-Dekker root bracketing method.  The initial estimated radius is calculated based on a crude density assumption for each layer (15, 5, 2, and $10^{-3}$ for iron, mantle, ice, and atmosphere respectively in the unit of \gcmc).

Predetermined by the user, the values of the enclosed mass at each layer interface are set as the bounds of the ODE integrator.  In contrast, the enclosed mass where the phase changes within a layer is determined by the phase diagram in $P$-$T$ space.  Thus, the step of the ODE integral typically does not land exactly on the location of the phase change.  To avoid introducing extra inaccuracy, when the ODE integration reaches a different phase the integrator is restored to the previous step and the integration step is cut into half.  The integrator can only move forward to the new phase when the step size that crosses the phase boundary is less than the ODE integrator accuracy tolerance multiplied by $M_p$.

Using the estimated boundary condition, we conduct the shooting to a fitting point method, which start ODE integration from both $m=0$ and $m=M_p$ toward a fitting point with mass $m_{fit}=0.2 M_p$.  The code will automatically adjusted $m_{fit}$ if it occurs at a layer's boundary.  The code uses GSL's  $gsl\_multiroot\_fsolver\_hybrids$ to adjust $P(0)$, $T(0)$, and $R_p$, until $P$, $T$, and $r$ at $m_{fit}$ obtained integrating from the inner branch and from the outer branch agrees within a relative error $<$10$^{-4}$.  The ODE integrator tolerance will be reduced (made more strict) by a factor of three if the multidimensional root-finder can not find a solution within 15 tries.  After the ODE integrator tolerance is reduced four times (three additional times, each by the stated factor of three), if the solver still cannot find a solution that satisfies the required joint accuracy at the fitting point, the code will output an error message as well as the best result it can find.  

The solver returns a planet interior profile object after completion, which includes the $r$, $P$, $T$, $m$, $\rho$, and the component phase at each grid step.  A {\it print} function is available to save the profile into an ASCII table.  The function {\it getRs} returns the list of radii of the outer boundary of each component.  Users can use {\it getstatus} to access the status information of the returned object which indicates whether precautions should be taken regarding the iteration result.  A list of potential problems and their assigned return value is summarized in Table~\ref{tab:status}.

\begin{table}
  \centering
  \caption{List of status information of the planet profile object}
  \label{tab:status}
  \begin{tabular}{cl}
    \hline
    Value & Meaning \\
    \hline
    0 & Normal. \\
    1 & The result includes phase(s) that is(are) \\ 
      & not formally implemented in \magra. \\
    2 & The two shooting branches do not match  \\ 
      & within the required accuracy at the fitting point.  \\
    3 & Under two-layer mode, the error of planet surface \\ 
      & pressure is larger than the required accuracy.  \\
    \hline
  \end{tabular}
\end{table}

\subsubsection{Simplified two layer mode}
\label{sec:2layer}

In addition to the regular solver, we include a mode for a simplified two layer solver.  We use this for quick calculations of water/mantle and mantle/core planets in \citet{Huang21}.  We recommend using our complete solver in most instances, but this method can be used for quicker and cheaper calculations; one such use is demonstrated in Sec.~\ref{sec:eosunc}.

Two-layer mode can only calculate the structure of 300 K isothermal planets and does not support an atmosphere.  Without the temperature differential equation, the program only solves the radius and pressure differential equations, with the radius boundary condition at the center, and the pressure boundary condition at the surface.  This simpler problem is solved using the "pure" shooting method inside-out, starting from the equations' singular point at the planet center.  The solver iterates the center pressure using the Brent-Dekker root bracket method until either the center pressure reaches the set accuracy target, or the surface pressure $P(M_p)$ is within 2\% of the user specified value.  If the iteration ends because the first criteria is satisfied first, an abnormal status value 3 will be returned by {\it getstatus}.  To avoid this, users may adjust the ODE integration tolerance and center pressure accuracy target accordingly based on the specific problem.

With this solver, the two-layer input modes provide an interface to calculate mass-radius curves for hypothetical planets that are composed of only two components with fixed mass ratio (e.g. figure 3 in \citet{Huang21}).   

\subsection{Phase diagram implementation}
\label{sec:phase}

Phase diagrams are defined in \emph{phase.cpp}.  The file contains four functions corresponding to the four layers: \emph{find\_Fe\_phase}, \emph{find\_Si\_phase}, \emph{find\_water\_phase}, and \emph{find\_gas\_phase}.  The \emph{if} statement is used within each function to create the region of P-T space to which an EOS applies.  The return value of each \emph{if} statement should be the name of the pointer corresponding to the EOS for the given phase (further described in Appendix~\ref{app:eos-struct}).

Transitions can be defined as pressure or temperature conditionals.  Our default phase diagrams are shown in Fig.~\ref{fig:phasediagram}.  We use phase transitions which are linear with pressure/temperature within the conditionals for the transfer between solid and liquid iron \citep{Dorogokupets17} and between bridgmanite and post-perovskite \citep{Ono05}.  Parameterized phase transition curves can also be defined as separate functions in \emph{phase.cpp} and called within the ``find phase'' functions.  We define the function \emph{dunaeva\_phase\_curve} to implement the fitting curve with five coefficients, $T(P) = a + bP + c \ln{P}+d/P+e\sqrt{P}$, from \citet{Dunaeva2010} for the transitions between phases of water.

\subsection{Built-in Equations of State}
\label{sec:eoslist}

We have over 30 EOSs available in \magra\ which can be called within the phase diagrams.  These include EOS functions for various planet building materials, and different parameter estimates for the same material from various works.  We discuss the equations currently available for each layer in the following four subsections.  New EOSs can be added to \textit{EOSlist.cpp} by following the storage structure described in Appendix \ref{app:eos-struct}.  In Fig.~\ref{fig:alleos} we show mass-radius relationships up to ten Earth-masses for planets of one layer with a large selection of our implemented equations. The index of equations in Fig.~\ref{fig:alleos} is also noted in the text in brackets.

In addition to the equations listed below, the program includes tabulated EOS of iron, silicate, and water from \citet{Seager2007} [CS, MS, WS], who combined empirical fits to experimental data at low pressure and Thomas-Fermi-Dirac theory at high pressure.

\begin{figure}
    \centering
    \includegraphics[width=\columnwidth]{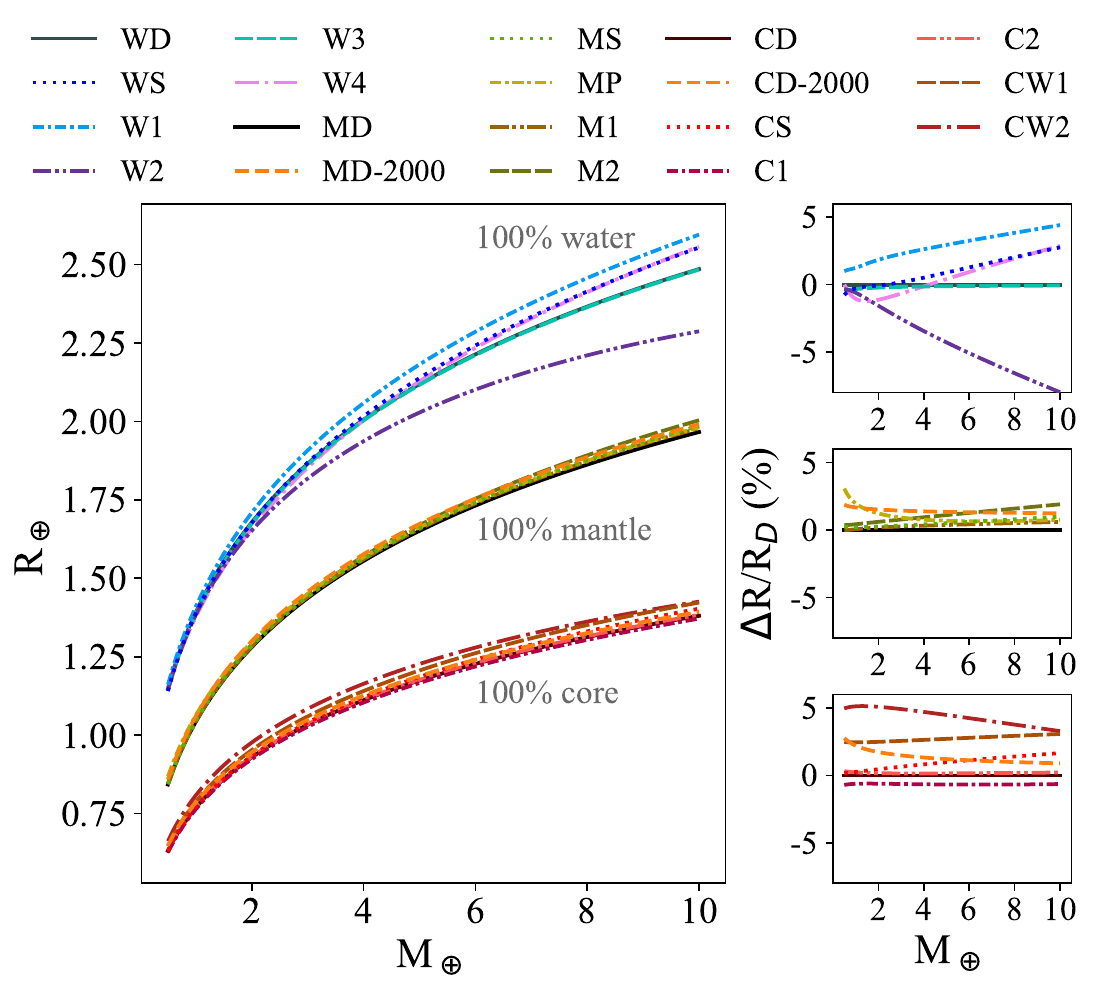}
    \setlength{\tabcolsep}{3pt}
    \begin{tabular}{lll}
            \hline
            Index &  Description & Source  \\
            \hline
            \hline
            *D & Default EoSs & Multiple\\ 
            *S & Tabulated &\citet{Seager2007}\\ 
            *-2000 & 2000 K surface & Multiple \\ 
            \hline
            \hline
            WD & Ice VI \& VII & \citet{Bezacier2014}\\
               & Ice X & \citet{Grande2022}\\ 
            W1 & Ice VII &\citet{Frank2004}\\ 
               & Ice X &\citet{French2009}\\
            W2 & Ice VII &\citet{Frank2004}\\ 
            W3 & Ice VII, VII', X & \citet{Grande2022}\\
            W4 & Ice VI \& VII & \citet{Bezacier2014}\\
               & Ice X & \citet{Hermann2011}\\
            \hline
            MD & Brg & \citet{Oganov04}\\
             & PPv & \citet{Sakai16}\\
            MP & PREM &\citet{Zeng2016}\\
            M1 & Brg/PPv &\citet{Oganov04}\\
            M2 & Brg &\citet{Shim2000}\\
            \hline
            CD & Fe HCP  &\citet{Smith2018}\\
            C1 & Fe HCP &\citet{Bouchet13}\\
            C2 & Fe HCP  &\citet{Dorogokupets17}\\
            CW1 & Fe-7wt\%Si &\citet{Wicks2018}\\
            CW2 & Fe-15wt\%Si &\citet{Wicks2018}\\
            \hline
        \end{tabular}
    \caption{\textit{Left}, mass-radius relationship for planets with 100 per cent of mass in either the core, mantle, or hydrosphere demonstrating many of the EoSs implemented in \magra. \textit{Right}, percent difference in final planet radii compared to our selected ``default'' EoSs for water (\textit{top}), mantle (\textit{middle}), and core (\textit{bottom}). Table lists the major components of each model with 300 K surface temperature unless designated with a ``-2000''. Near the surface, water planets have water and Ice VI, and hot mantle planets have silicate melt.
    }
    \label{fig:alleos}
\end{figure}

\subsubsection{Core/Iron}
\label{sec:iron}

At the extreme pressures of a planetary core, iron is stable in a hexagonal close-packed (HCP) phase.  The program includes HCP iron EOSs from \citet{Bouchet13} [C1], \citet{Dorogokupets17} [C2], and \citet{Smith2018} [CD] (see Fig.~\ref{fig:alleos}).  We choose as our default equation the Vinet fit (Eq.~\ref{eq:vinet}) from \citet{Smith2018} measured by ram compressing iron to 1.4 TPa.  We determine the fitting parameters of the \gru\ by fitting Eq.~\ref{eq:12} to Fig.~3b in \citet{Smith2018} through maximum likelihood estimation.  The default core layer includes a liquid iron EOS and melting curve from \citet{Dorogokupets17} who compressed iron to 250 GPa and 6000 K.

Iron-silicate alloy EOSs are implemented as well.  \citet{Wicks2018} experimentally determined the EOS for Fe-Si alloys with 7 wt per cent Si [CW1] and 15 wt per cent Si [CW2].  These alloy equations are useful in reproducing the density of the Earth's core which contains an unknown mixture of light elements \citep{earthcore}.  An additional liquid iron equation is included from \citet{Anderson94}.

At lower temperatures and pressures, fcc (face-centered cubic) iron is not currently implemented.  The EOS of fcc iron has similar parameters to that of hcp iron (K$_{0,fcc}=146.2$, K$_{0,hcp}=148.0$).  The fcc-hcp-liquid triple point is at 106.5 GPa and 3787 K \citep{Dorogokupets17}.

\subsubsection{Mantle/Silicate}
\label{sec:mantle}

The main mineral constituent of the Earth's mantle is bridgmanite (Brg), referred to in the literature cited here as silicate perovskite, which at high pressure transitions to a post-perovskite (PPv) phase \citep{Tschauner2020}.  \magra\ includes third-order BME (Eq.~\ref{eq:bme}) Brg from \citet{Shim2000} [M2] and Vinet EOSs (Eq.~\ref{eq:vinet}) for Brg and PPv measured with \textit{ab initio} simulations and confirmed with high pressure experiments from \citet{Oganov04} and \citet{Ono05} [M1].  

Our default mantle [MD] includes Brg from \citet{Oganov04} and an updated PPv thermal EOS from \citet{Sakai16}.  \citet{Sakai16} compressed PPv to 265 GPa with a laser-heated diamond anvil cell and extended the EOS using a Keane fit (Eq.~\ref{eq:keane}) to 1200 GPa and 5000 K with \textit{ab initio} calculations.  At high temperatures ($>$1950 K at  1.0 GPa), we use a liquid MgSiO$_3$ with RTPress EOS from \citet{Wolf2018} with melting curve from \citet{Belonoshko05}.  The silicate melt transfers directly to PPv when P$>$154 GPa and T$>$5880 K since the behavior in this regime is unknown. An alternate silicate melt is implemented from \citet{Mosenfelder2009}.

Our default mantle is thus pure MgSiO$_3$ in high-pressure phases which is more SiO$_2$ rich than the Earth.  At low pressure ($\lesssim$25 GPa), materials such as olivine, wadsleyite, and ringwoodite are the main constituents \citep{Sotin2007}.  Although we don't include these minerals and more complex mantle chemistry at this time, we include a tabulated EOS for the mantle using the mantle properties in the Preliminary Reference Earth Model (PREM) \citep[Appendix F]{BookPhyEarth} [MP].  We also have not explored additional measurements of transition curves between compositions/phases.  

\subsubsection{Hydrosphere/Water}
\label{sec:water}

Similar to the icy moons in the Solar System; exoplanets with low density are theorized to have a large fraction of mass in a hydrosphere composed of primarily high-pressure water-ice.  Laboratory measurements of the thermal properties of ice at high pressures are difficult to obtain and sometimes inconsistent \citep{Thomas2016,Myint2017}.  Due to the complexity of the phase diagram of water and large uncertainties in current measurements, in the current version of \magra, we use a simplified hydrosphere phase diagram.

At pressures below 2.216~GPa, we applied liquid water \citep{Valencia2007} if the temperature is above the melting curve \citep{Dunaeva2010}.  Below the melting curve, we have Ice Ih \citep{Feistel2006, Acuna2021} at pressures below 0.208~GPa and Ice VI \citep{Bezacier2014} (described below) at pressures above 0.632~GPa.  Ice II, III, and V, which exist at pressures in between, are currently not built-in since this layer is thin and has negligible impact on the planet.  If a planet passes through these phases, the Ice VI EOS is used  and the user is notified that they passed through this region.  For water and Ice Ih the isothermal temperature is applied regardless the value of the \textit{isothermal} flag.

At high pressure, our default hydrosphere layer uses Ice VII with thermal expansion \citep{Bezacier2014}.  Instead of using the MGD representation, most high pressure ice studies express their result with thermal expansivity, with the exception of \citet{Fei1993} which present their result with both methods.  Although the isobaric specific heat capacity $c_p$ is necessary to determine the temperature gradient using the thermal expansivity representation (see Eq.~\ref{eq:dTdm_alpha}), it is not reported in most studies.  Similar to \citet{Zeng2021}, we include a formulation using the thermal expansion representation for Ice VI and VII from \citet{Bezacier2014} with estimated $c_p$ of 2.6 and 2.3~$\rm J\,g^{-1}\,K^{-1}$ based on \citet{Tchijov2004}.

At this time, we do not include a phase boundary or EOS for supercritical water.  To avoid extrapolating the Ice VII EOS to high temperature that may lead to non-physical density, an artificial melting curve at constant temperature 1200~K is drawn in between 2.216~GPa and 30.9~GPa.  The water EOS \citep{Valencia2007} is used as an approximation in supercritical regions and the user is notified.  Lastly, the default hydrosphere transitions at 30.9 GPa to a Vinet EOS for Ice X from \citet{Grande2022} which is calculated at 300 K and returns a constant temperature throughout regardless the value of the \textit{isothermal} flag.  

The following high-pressure water-ice EOSs are also available as alternative. Third-order BME Ice VII EOS with thermal expansion and heat capacity from \citet{Frank2004} and 3rd order BME Ice VII EOS using MGD expression from \citet{Sotin2007}.  Other ice EOSs that implemented do not support the isentropic temperature gradient.  They include Ice VII EOS along the melting curve \citep{Frank2004} and tabulated Ice X \citep{French2009} which combined were used in the planetary models of \citet{Zeng2016} [W1].  Our library has three fits to the 300 K Ice VII measurements from \citet{Frank2004} -- their original 3rd order BME, their BME parameters in a Vinet equation [W2], and a Vinet fit to their results.  We have the three phases of ice including the transitional Ice VII' from \citet{Grande2022} which was used in \citet{Huang21} [W3].  Lastly, we have a 3rd order BME Ice X from \citet{Hermann2011} [W4].  A future version of \magra\ will expand upon these options with results from \citet{French2015}, \citet{Journaux2020}, and other water phases from the AQUA collection \citep{Haldemann2020}.

\subsubsection{Atmosphere/Gas}
\label{sec:gas}

The ideal gas equation of state:
\begin{equation}
  \label{eq:ig}
  P=\frac{\rho R T}{m_{mol}},
\end{equation}
and temperature relation:
\begin{equation}
    \label{eq:Tgas}
    T \propto P^{(\gamma_{\rm gas}-1)/\gamma_{\rm gas}}
\end{equation}
is applied to the gas layer at all pressures and temperatures, where $\gamma_{\rm gas}$ is the adiabatic index of the gas.  Combining Eq.~\ref{eq:1}, \ref{eq:2}, \ref{eq:ig}, and \ref{eq:Tgas}, we have the ideal gas temperature gradient
\begin{equation}
    \label{eq:dTdm_gas}
    \frac{\dd T(m)}{\dd m} = -\frac{(\gamma_{\rm gas}-1)m_{mol}Gm}{4\pi r^4\gamma_{\rm gas}R\rho}.
\end{equation}

The temperature profile can be chosen between isothermal and isentropic by setting the number of atoms per molecule $n$ (see Table~\ref{tab:param}), which further determines $\gamma_{\rm gas}$.  The adiabatic index $\gamma_{\rm gas}$ can be chosen from $\gamma_{\rm gas}=\frac{5}{3}$ for monatomic gas, $\gamma_{\rm gas}=\frac{7}{5}$ for diatomic gas, and $\gamma_{\rm gas}=\frac{4}{3}$ for polyatomic gas.  An isothermal gas EOS can be achieved by setting its $n=0$.  The temperature profile of the gas layer is determined by the properties of the gas EOS and will not be overwritten by the {\it isothermal} flag. 

\citet{Nixon2021} shows that self-consistently modelled atmosphere temperature profiles can be closely matched by a radiative profile above a radiative-convective boundary, and an adiabatic profile below the boundary.  The pressure of the radiative-convective boundary can span from 1 bar to over 1 kbar, mainly depending on the intrinsic effective temperature and equilibrium temperature of the planet \citep{Fortney2007}.  To simulate this temperature structure, the default settings for the atmosphere in \magra\ include an isothermal atmosphere at $P<100$~bar to approximate the radiative temperature gradient, and an adiabatic temperature gradient at $P\ge 100$~bar.  The mean molecular weight of each gas EOS is fixed.  The dependence of planetary radius on atmosphere mean molecular weight is explored in Fig.~\ref{fig:ternatm}.  

Radiative temperature profile and coupled atmosphere-interior models such as \citet{Mousis2020} are potential future avenues of work.  Additionally, we anticipate that future versions of \magra\ will include non-ideal atmospheres from \citet{Saumon1995} and \citet{Chabrier2019}.

\section{Known Limitations}
\label{sec:limitations}

We design \magra\ with a focus on extensibility.  \magra\ permits users to extrapolate EOS functions beyond the temperature and pressure bounds from experimental and theoretical works.  The phase diagram can be set up to allow a material where it would not be physical (e.g. use water in place of supercritical water).  If the integrator passes through these extrapolated regions, it may find an incorrect solution or may fail to continue to integrate.  Here we detail some limitations one should be aware of for each layer with our current defaults.

For the core, the liquid iron EOS from \citet{Dorogokupets17} has no solution when the temperature of a given step gives a thermal pressure much larger than the total pressure at that step.  This occurs most often when the iron core outer boundary is near the surface of the planet and at unusually high temperatures.  In this case, a \textit{no solution found} tag will be returned.

Extrapolated EOS functions may have two density solutions that satisfy the equation set at given $P$ and $T$.  The nonphysical solution typically has a lower density and the $\frac{\partial \rho}{\partial P}<0$.  The Newton's method density solver may converge to the incorrect branch at a transition from a low density phase to a higher density phase because the initial density guess of the high density phase is too small.  We experienced this problem in the mantle when transitioning from Si melt to Brg from \citet{Oganov04} at low pressures.  To avoid this, if the initial density guess gives $\frac{\partial \rho}{\partial P}<0$, a density larger than $\rho(P=0, T=T_0)$ will be applied as the initial density guess of the Newton's solver, instead of the density solution of the previous step.

For the hydrosphere, users should be aware that \magra\ will use the water EOS and return a constant temperature even if the water would be in the vapor or supercritical phase.  Because water EOS is softer and the thermal expansion is not included, the density is overestimated in supercritical regions.  This may lead to a density drop at the phase transition from supercritical fluid to Ice X.  Other phases without thermal parameters, such as Ice X, will return a constant temperature.  Lastly, the adiabatic portion of the default atmosphere quickly rises in temperature and pressure and could force internal layers into extreme regions of the phase diagram.  Ideal gas atmospheres with constant mean molecular weight are less realistic for large gas envelopes with pressures over 0.1 GPa \citep{Zeng2021, Nixon2021}.

\section{Test problems and utility}
\label{sec:test-problems}

\subsection{One-Earth mass and comparison with ExoPlex}
\label{sec:exoplex}

\begin{figure}
    \centering
    \includegraphics[width=\columnwidth]{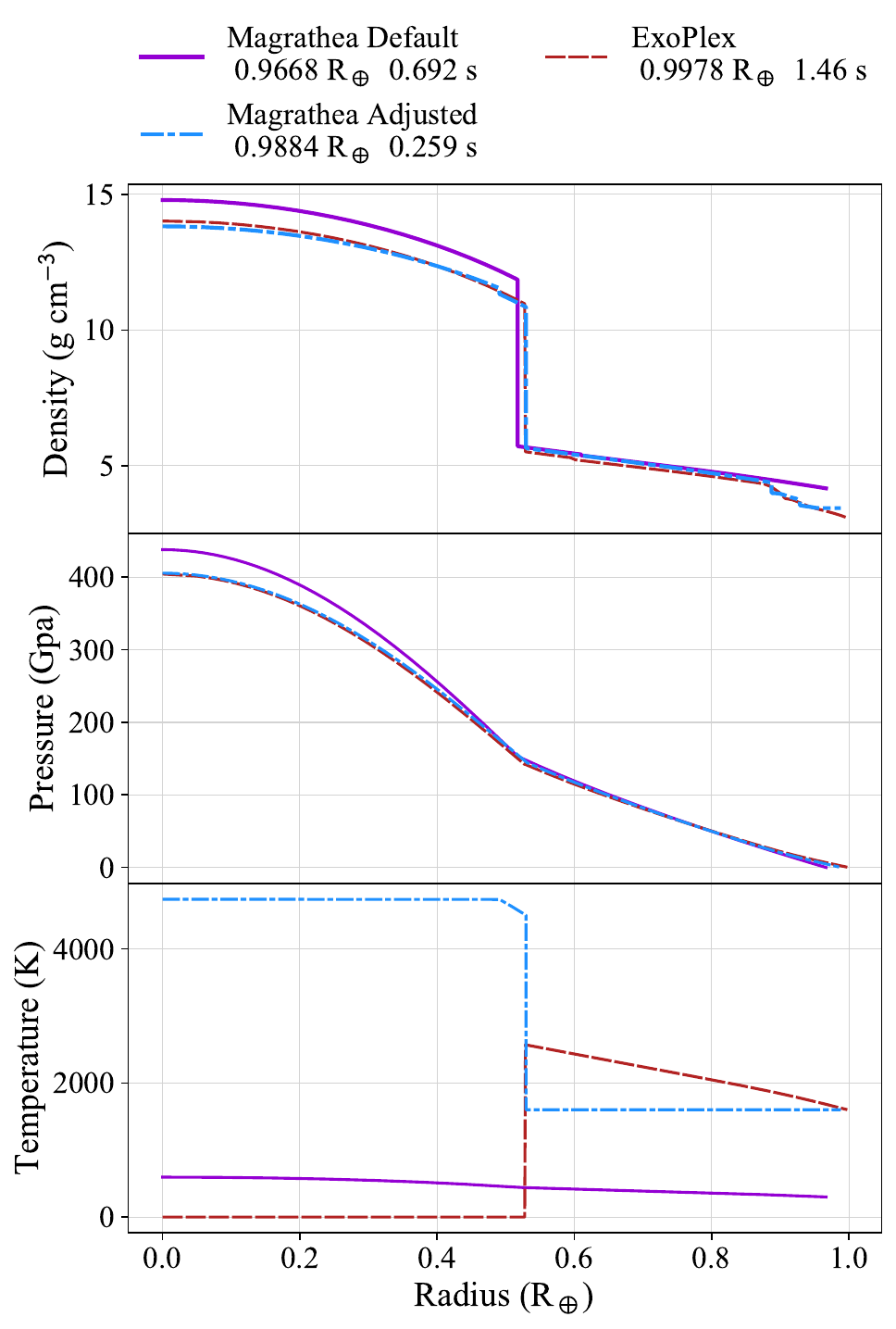}
    \caption{Density, pressure, and temperature verses radius solution for a two-layer, one Earth-mass planet with 33 per cent by mass core. The three models shown are \textit{Magrathea Default} with hcp-Fe core \citep{Smith2018} and Brg/PPv mantle \citep{Oganov04,Sakai16}, \textit{Magrathea Adjusted} with Fe-Si alloy core \citep{Wicks2018} and PREM mantle \citep{Zeng2016} and a temperature discontinuity, and the default settings in \textit{ExoPlex}. \textit{Magrathea Default} is set to 300 K at the surface while \textit{ExoPlex} suggests a 1600 K mantle. Temperature is solved throughout \textit{Magrathea Default} and in \textit{ExoPlex}'s mantle. Magrathea returns no change in temperature when a phase does not have temperature parameters available while ExoPlex returns zero Kelvin. The planet's radius and the average run time over 100 integrations is listed in the legend.}
    \label{fig:compare}
\end{figure}

We turn our attention to showing the outputs and utility of \magra.  We first simulate a one-Earth mass, two-layer planet with our full solver.  The planet has a structure similar to Earth with 33 per cent of its mass in the core \citep{BookPhyEarth}.  With the default EOSs described in Sec.~\ref{sec:eoslist} and a surface temperature of 300 K, \magra\ produces a planet with radius of 6166 km or 0.967 R$_\oplus$.  Fig.~\ref{fig:compare} shows the pressure, density, and temperature found throughout the planet.

Our default model differs from a detailed model of Earth in that our mantle is only Brg/PPv, the core has no lighter elements, and there is thermal equilibrium between the layers.  Rather than choosing our default settings, users can choose to use an iron-silicate alloy EOS \citep{Wicks2018} in the core and PREM in the mantle.  The user can also start the mantle hot at 1600 K and implement a 2900 K jump across the core-mantle boundary which creates a layer of liquid iron.  This ``adjusted'' model, shown in Fig.~\ref{fig:compare}, has a radius within 1.16 per cent of the radius of Earth (0.9884 R$_\oplus$). 

We compare our Earth-like planets to one created with a version of \textit{ExoPlex}\footnote{\url{https://github.com/CaymanUnterborn/ExoPlex}}, an open-source interior structure solver written in Python.  \textit{ExoPlex} is used in \citet{Unterborn2018,Unterborn2019,Schulze2020}.  The code uses a liquid iron core \citep{Anderson94} and self-consistently calculates mantle phases with \textit{Perple\_X} \citep{Connolly2009} using EOS for mantle materials from \citet{Stixrude2005}.  \textit{ExoPlex} comes with predetermined grids of mineralogy to capture the mantle chemistry of planets between 0.1 and $\sim$2 M$_\oplus$. 

As seen in the top panel of Fig.~\ref{fig:compare}, \textit{ExoPlex} captures changes in density in the upper mantle that our Brg/PPv model does not.  \textit{ExoPlex} creates an Earth-like planet with a radius of 0.9978 $R_\oplus$ which is 3.2 per cent larger than our default planet and 1.0 per cent larger than our adjusted model.  At least half of this difference is from a different choice in core EOS; \textit{ExoPlex}'s innermost density is 14.0 while our default is 14.8 \gcmc.  The rest of the difference is \textit{ExoPlex}'s ability to capture upper-mantle chemistry which decreases the density from our default for approximately 15\% of the planet's mass.

\subsection{Run Time}
\label{sec:timing}

\begin{figure}
    \centering
    \includegraphics[width=\columnwidth]{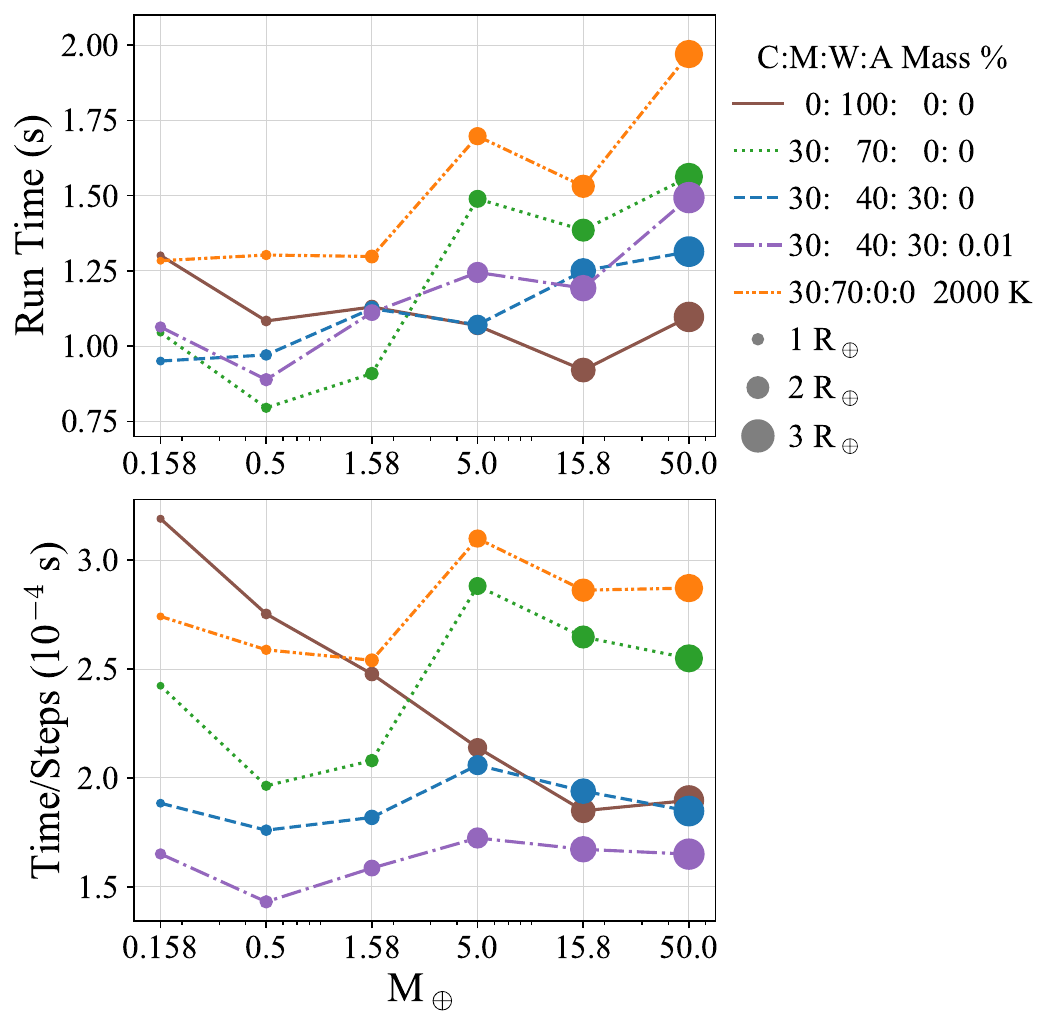}
    \caption{Plot of run time, \textit{top}, and of run time divided total number of steps, \textit{bottom}, for our default model across six planet masses and four compositions: 100\% core, 30\% core 70\% mantle, 30\% core 40\% mantle 30\% water, and 30\% core 40\% mantle 29.99\% water 0.01\% H/He atmosphere with 300 K surface temperature.  The last set of planets, \textit{orange}, have 30\% core, 70\% mantle, and surface temperature of 2000 K.  The run time is the average time measured across 100 runs.  Sizes of the markers are proportional to the planet radius. }
    \label{fig:time}
\end{figure}

We perform timing tests for both \magra\ and \textit{ExoPlex} for 100 repeat simulations of the Earth-like planet from the above section.  \magra\ takes on average 0.69 s to converge.  This is 2.1 times faster than \textit{ExoPlex} (1.46 s).  The ``adjusted'' model uses a tabulated EOS for the mantle and a core EOS with no thermal information, thus most of the planet is isothermal with temperature only solved in the small region of liquid iron.  This model runs 5.6 times faster than \textit{ExoPlex}.  Average times are also listed in Fig.~\ref{fig:compare}.

The run time of \magra\ is dependent on the mass of the planet, the user's choice of planet model, and the error tolerances.  In Fig.~\ref{fig:time}, we show run times of some typical planet masses and compositions with our default settings.  Four compositions are run with a surface temperature of 300 K.  A final set has a surface temperature of 2000 K which has a region of silicate melt at the surface.  The figure shows that time remains near one second, but is not easily determined by input mass and composition.  Here the average run time is 1.2 seconds.  The longest and shortest run times are a factor of 2.5 times slower/faster.  These solutions take between 15 and 30 iterations, and the final solutions have between 520 and 840 steps in mass.

In general the run time increases with mass, though not strictly since the step size is adaptable.  Crossing phase or compositional boundaries costs an insignificant amount of run time.  However, certain phases may take shorter or longer to solve.  The planets with atmospheres have the most compositional layers, but the ideal gas solves quickly resulting in the shortest time per step.  The runs which take the longest to converge are large, hot planets with a surface of silicate melt and large mantles. 

\subsection{Uncertainty from equation of state}
\label{sec:eosunc}

\begin{table*}
  \centering
  \caption{EOS parameters for \citet{Smith2018}, \citet{Sakai16}, and \citet{Grande2022} with uncertainty and the uncertainty in radius for a 10 M$_\oplus$ single-layer planet from uncertainty in EOS parameters.}
  \label{tab:eosunc}
  \addtolength{\tabcolsep}{3pt}
  \begin{tabular}{cccccc}
        \hline
        Phase & TP & $V_0$ & $K_0$ & $K_0'$& $\sigma_R/\mu_R$ for 10 M$_\oplus$ \\
         & GPa & cm$^3$~mol$^{-1}$ & GPa & GPa& \%\\
        \hline
        \hline
        Fe HCP& - & 6.625 & 177.7(6) & 5.64(1)& 0.089\\
        \hline
        PPv & 112.5(8.1) & 24.73 & 203(2) & 5.35(9)&0.19 \\
        \hline
        ice-VII & - & 12.80(26) & 18.47(4.00) & 2.51(1.51)& \\
        ice-VII$\rm_t$ & 5.10(50) & 12.38(16) & 20.76(2.46) & 4.49(35)& 0.24\\
        ice-X & 30.9(2.9) & 10.18(13) & 50.52(16) & 4.5(2) &\\
        \hline
  \end{tabular}
\end{table*}

\magra's structure allows the user to change the EOS and test their effects on planet structure.  In Fig.~\ref{fig:alleos}, we show how choice of EOS can change the radius of single layer planets of masses from one to ten Earth-masses.  For the hydrosphere, further discussed in \citet{Huang21}, new measurements of Ice VII, Ice VIIt, and Ice X change the resulting planet radius from those in \citet{Zeng2016} by 1-5 per cent across this range of masses.  

The mantle's EOS has the smallest effect on radius in agreement with \citet{Unterborn2019}.  Comparing the tabulated EOS from PREM to our default mantle, we find that the radii of pure mantle planets differ by less than one percent for 2-5 M$_\oplus$.  Planets $<$2 M$_\oplus$ have a larger difference in simulated radii.  Capturing the density of the upper mantle is more important for these smaller planets.  

HCP iron measurements generally agree when looking at bulk planet properties.  However, it is uncertain how much lighter elements can be incorporated into a planet's core.  Included in \magra\ are two silicate alloy EOSs from \citet{Wicks2018}.  Using a 15 per cent by weight silicate alloy EOS results in a 1 M$_\oplus$, core-only planet that differs by five per cent in radius to our default.  For comparison to the above radius changes, the 1-sigma uncertainties in radii of the seven Trappist-1 planets are 1-2 per cent \citep{Agol2020}.

A single measured EOS also has uncertainties in its own parameters.  \magra\ includes the ability to modify an EOS at run-time to determine the effects that measurement uncertainty has on a planet radius estimates.  This feature is available in \textit{input\_mode} 3, 4, and 5.  It's implementation is further detailed in Appendix \ref{app:modify-eos}.  The first two modes are limited to using the isothermal \textit{twolayer} solver explained in Sect. \ref{sec:2layer}.  The \textit{input\_mode=5} runs the full model with user specified planet composition and mass.  \magra\ iterates over EOS parameters from an input file and outputs the planet radius for each EOS modification.

In Table \ref{tab:eosunc}, we show the reported uncertainties in three experimental EOS from \citet{Sakai16,Smith2018,Grande2022}.  We model 10 M$_\oplus$ planets each with a single layer.  We then modify the EOS of that layer from 1000 values of the reference volume, bulk modulus, and the derivative of the bulk modulus drawn from Gaussians centered at the reported mean and using the given uncertainties.  For the 100 per cent mantle and 100 per cent hydrosphere models we also draw 1000 values of the transition pressure (TP) between phases.  The mantle planets have a Brg phase, but it's parameters are held constant as they are derived from \textit{ab initio} simulations.  The resulting uncertainty in planet radius for each single-layer model is $<1$ per cent for 10 M$_\oplus$ planets.  Although the choice of EOS is of more consequence to a single measurement's uncertainty, this module allows users to investigate new measurements and their associated uncertainties.

\subsection{Ternary diagram}\label{sec:ternary}

\begin{figure*}
    \centering
    \includegraphics[width=\textwidth]{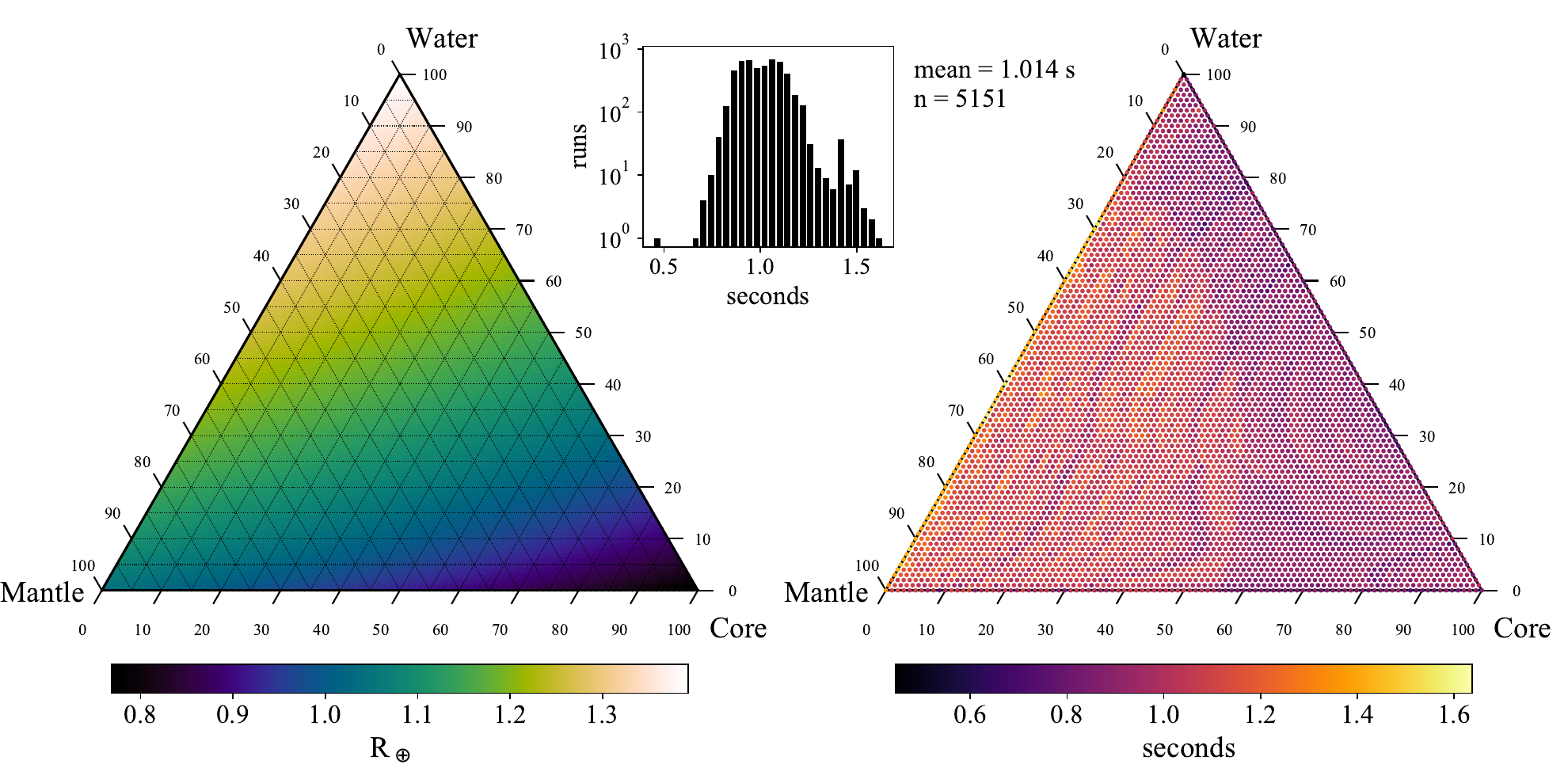}
    \caption{\textit{Left}, ternary diagram where the axes are the percentage of mass in a core, mantle, and water/ice layer.  The radius in Earth radii is shown by the color scale for 5151, one Earth-mass planets at integer percentages with our default model.  Color map is interpolated between the simulations.  \textit{Right}, plot of \magra's run time for each planet. \textit{Middle}, histogram of run time on a log-log scale showing mean time of 1.014 seconds.  Ternary plots are generated with the python-ternary package by \citet{ternaryplot} with colormaps from \citet{cmasher}.}
    \label{fig:tern}
\end{figure*}

In Section \ref{sec:eoslist}, we show mass-radius relationships for single-layer planets.  However, these relationships are not unique when considering a planet with three or more layers.  For a three-layer planet, ternary diagrams provide a way to visualize the radius parameter space for a planet with a certain mass (conversely we could show the various masses for a certain radius).  Ternary diagrams have been used for exoplanet interiors in \citet{Valencia2007tern}, \citet{Rogers2010}, \citet{Madhusudhan2012}, \citet{Brugger2016}, \citet{Neil2020}, and \citet{Acuna2021}.

The location of a point on the equilateral triangle is given by the percentages of mass in the three layers---core, mantle, and water.  The three coordinates add to a constant 100\%.  Since axes are interchangeable, we decide to use the same representation as \citet{Zeng2008}.  With this orientation the core to mantle ratio increases along the bottom axes and the water percentage increases perpendicular to the bottom axis. 

We include with \magra, a ternary plotting python script that uses python-ternary by \citet{ternaryplot}.  To run 5151 planets with integer percentages of mass in each layer we use \magra's bulk input mode (\textit{input\_mode=6}).  This function requires an input file that contains each planet's total mass and the mass fraction of core, mantle, and hydrosphere.  Any mass not allocated to a layer is put into an atmosphere layer.  

In Fig.~\ref{fig:tern} \textit{left} each position on the triangle gives the radius of a unique one Earth-mass planet.  We use planets in thermal equilibrium, a 300 K surface temperature, and with our default EOSs.  This model underestimates the Earth's radius as discussed in Sec.~\ref{sec:exoplex}.  A dry, one Earth-mass, and one Earth-radius planet in these simulations has approximately a 19 per cent core mass fraction (CMF).  Because of the difference between water's and iron's density, we can see that the radius varies most dramatically along the water-core axis.  A one Earth-mass and Earth-radius planet is also consistent with having 75 per cent CMF and 25 per cent of mass in a hydrosphere.

The entire suite of simulations needed for the one Earth-mass ternary takes approximately 87 minutes to complete.  The time for each run is shown in Fig.~\ref{fig:tern}, \textit{right}.  The average run time is 1.014 seconds.  In agreement with Fig.~\ref{fig:time}, time has a weak correlation with mantle percentage.

\subsubsection{Ternary with atmosphere}

\begin{figure}
    \centering
    \includegraphics[width=\columnwidth]{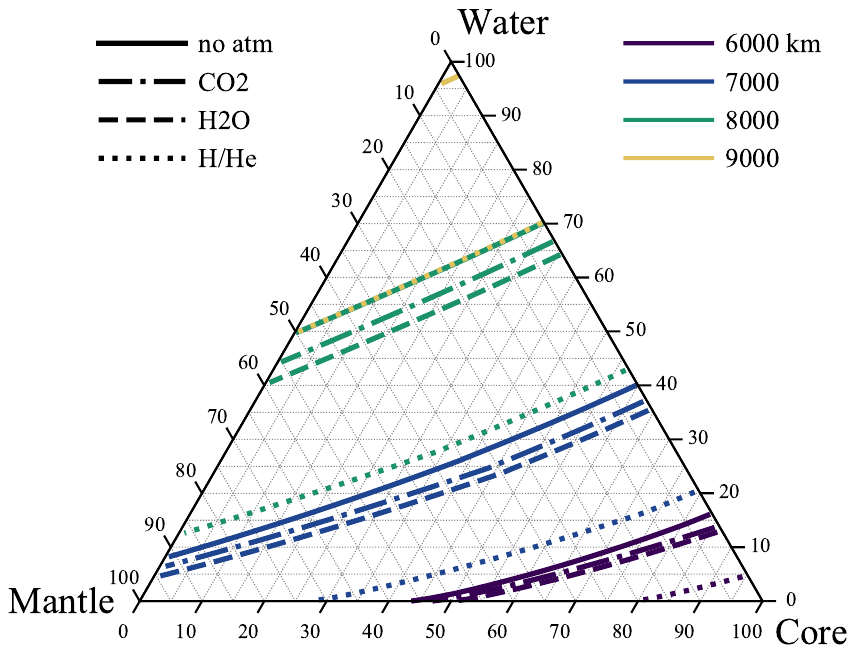}
    \caption{Core-Mantle-Water mass percentage ternary plot with colored contours of constant radius for one Earth-mass planets.  Four types of planets, represented by the line-style, are calculated: one with no atmosphere and three with 0.01 per cent of their mass in an atmosphere layer. The atmosphere mass was subtracted equally from both the mantle and core to keep total mass equal to one Earth-mass. The three atmospheres have varying mean molecular weight: CO$_{2}$ with 44 g~mol$^{-1}$, H$_{2}$O with 18 g~mol$^{-1}$, and H/He mixture with 3 g~mol$^{-1}$.}
    \label{fig:ternatm}
\end{figure}

\label{sect:atmgraph}
As a final example, we add an atmosphere to our ternary planets using our default two ``phase'' ideal gas atmosphere.  In Fig.~\ref{fig:ternatm}, we show contour lines of constant radius.  The solid lines come from Fig.~\ref{fig:tern}.  The 5151, one Earth-mass planets are simulated again with 0.01 per cent of their mass taken equally from the core and mantle and put it into an atmosphere layer.  This is comparable to the atmospheric mass fraction of Venus.  Three types of ideal gas are used with varying molecular weight.  

A small atmosphere of a hydrogen/helium mixture inflates the planets by almost 1000 km for all interior structures.  A 6000 km, one-earth mass planet with a small H/He atmosphere is consistent with a 2-layer planet of 20:80 mantle:core (intersection of mantle-core axis, bottom) or 5:95 water:core (intersection of core-water axis).  Together, the one Earth-mass ternaries explore the radii of a large range of possible interior (plus atmosphere) structures.

\section{Summary}
\label{sec:summary}

In this work, we presented \magra, an open-source planet interior solver.  \magra\ is available at \url{https://github.com/Huang-CL/Magrathea}.  Given the mass in the core, mantle, hydrosphere, and atmosphere, the code uses a shooting method to find the planet's radius and internal conditions. 

The code allows users to easily modify their planet models.  Each layer is given a phase diagram where the EOS may change based on P-T conditions.  Multiple formulations, both thermal and non-thermal, for the EOS are implemented.  We document in this work our built-in library of EOSs and how to add new equations.  Our default EOSs feature up-to-date experimental results from high-pressure physics.  We show that the choice of planet model has an effect on inferences of planet composition.

The timings and results presented here show that \magra\ is an efficient and useful tool for characterizing the possible interiors of planets.  \magra\ is currently under active development and we plan future expansions to the package.  In this version, the isentropic temperature gradient is only available for the Ice VI and Ice VII in the hydrosphere.  Future work includes implementing additional core alloys, upper mantle materials, more phases of water, a thermodynamic Ice X, atmosphere EOSs, and a graphically determined phase diagram.  We are currently working on a planet composition finder which returns a sample of possible interiors from a given radius and mass.  We encourage the community to contribute and use \magra\ for their interior modeling needs.  

\section*{Acknowledgements}

CH acknowledge support by the NASA Exoplanet Research Program grant 80NSSC18K0569. DRR and JHS thank the National Science Foundation for their support for our work under grant AST-1910955.  We thank Ashkan Salamat and Oliver Tschauner for their important help in creating the EOS list. DRR thanks Cayman Unterborn for the use of ExoPlex. 

\section*{Data Availability}
The data underlying this article will be shared on request to the corresponding authors, or can be accessed online at our GitHub repository.

\bibliographystyle{mnras}
\bibliography{unlvplanet}

\appendix
\section{EOS storage structure}
\label{app:eos-struct}

Calculating the density for a variety EOS formulae or tabulated EOS is packaged in the structure \emph{EOS} declared in \emph{EOS.h} and accomplished in \emph{EOS.cpp}.  The EOS of each phases needed for the calculation is an object of the \emph{EOS} structure.

The name of pointers to \emph{EOS} objects should be declared as an external variable in the header file \emph{EOSlist.h} and then accomplished in \emph{EOSlist.cpp}.  To use the formula method, user should create a dictionary style two-dimensional double array with a shape of (\emph{length},2), where \emph{length} is the number of provided EOS parameters.  Table~\ref{tab:formula} - \ref{tab:param} explains all acceptable parameters.  In the array, the first number of each row is the index key of the parameter according to Table~\ref{tab:param}.  The second number is the value in the required unit.  Not all spaces in the table need to be filled up.  Not available values can be skipped.  $V_0$ and $m_{mol}$ are required for all types of EOSs.  $K_0$ is required for condensed phase material.  $\gamma_0$ and $\beta$ are required for isentropic temperature profile and thermal expansion.  The minimum required parameters can be determined by the corresponding formulae listed in Section~\ref{sec:formula} and default values in Table~\ref{tab:param}.

\begin{table}
\begin{threeparttable}
  \centering
  \caption{Index for types of isothermal EOS formulae}
  \label{tab:formula}
  \begin{tabular}{ccl}
    \hline
    Index & Function Name & Comment \\
    \hline
    0 & BM3 & 3rd order Birch-Murnaghan\tnote{a}\\
    1 & BM4 & 4th order Birch-Murnaghan \\
    2 & Vinet &  \\
    3 & Holzapfel & \\
    4 & Keane & \\
    5 &  & Empty \\
    6 &  & Ideal gas law \\
    7 &  & Interpolation \\
    8-12 & Same as 0-4 &  In combination with RTPress \\
    \hline
  \end{tabular}
  \begin{tablenotes}
    \item[a] Default
  \end{tablenotes}
  \end{threeparttable}
\end{table}

\begin{table}
\begin{threeparttable}
  \centering
  \caption{Temperature profile option \tnote{a}}
  \label{tab:temp}
  \begin{tabular}{cl}
    \hline
    Index & Temperature profile calculation method \\
    \hline
    0 & No temperature profile, must be isothermal\\
    1 & External entropy function \\
    2 & External temperature gradient function \tnote{b} \\
    3 & Ideal gas \\
    4 & EOS is fitted along isentrope \tnote{c} \\
    5-7 & Isentropic curve \tnote{d}\\
    8 & RTPress  \\
    9 & Thermal expansion \tnote{e}\\
    \hline
  \end{tabular}
  \begin{tablenotes}
    \item[a] Only need to specify this option in the input list if using external entropy function or external temperature gradient function (index 1 or 2), or the EOS is fitted along isentrope (index 4).  Otherwise, the code will determine the option based on the parameters provided.
    \item[b] The only method to set the gradient is using the modify\_extern\_dTdP function.
    \item[c] Selecting this option requires parameters to calculate Debye thermal pressure in Eq.~\ref{eq:14}. The isentrope temperature profile is enforced to the phase with this option even the isothermal option is chosen for the planet solver.
    \item[d] Selecting this option requires parameters to calculate Debye thermal pressure in Eq.~\ref{eq:14}.
    \item[e] Selecting this option requires $\alpha$ and $c_p$.
  \end{tablenotes}    
\end{threeparttable}
\end{table}

\begin{table*}
  \centering
  \caption{List of EOS parameters }
  \label{tab:param}
  \begin{threeparttable}
  \begin{tabular}{cccl}
    \hline
    Index & Variable & Unit & Comment \\
    \hline
    0 & EOS formula type & & See table~\ref{tab:formula} \\
    1 &	$V_0$ & cm$^3$~mol$^{-1}$ & Molar volume at reference point \\
    2 &	$K_0$ & GPa & Bulk modulus \\
    3 &	$K_0'$ & & Pressure derivative of the bulk modulus. Default 4 \\
    4 &	$K_0''$ & GPa$^{-1}$ & Second pressure derivative \\
    5 &	$m_{mol}$ & g~mol$^{-1}$ & Molar mass \\
    6 &	$P_0$ & GPa & The minimum pressure, corresponding to $V_0$. Default 0  \\
    7 &	$\Theta_0$ &K& Fitting parameter of Einstein or Debye temperature (see Eg.~\ref{eq:13}). Default 1 \\
    8 &	$\gamma_0$ & & Fitting parameter of Gr\"{u}neisen parameter (see Eq.~\ref{eq:12} and \ref{eq:RTpress_gru})\\
    9 & $\beta$ & & Fitting parameter of \gru\ parameter (see Eq.~\ref{eq:12} and \ref{eq:fT})\\
    10 & $\gamma_\infty$ & & Fitting parameter of \gru\ parameter (see Eq.~\ref{eq:13}). Default 2/3 \\
    11 & $\gamma_0'$ & & Volume derivative of the \gru\ parameter (see Eq.~\ref{eq:RTpress_gru}) \\
    12 & $e_0$ & $10^{-6}$ K$^{-1}$ & Electronic contribution to Helmholtz free energy (see Eq.~\ref{eq:9}). Default 0 \\
    13 & $g$ & & Electronic analogue of the \gru\ parameter (see Eq.~\ref{eq:9}) \\
    14 & $n$ & & Number of atoms in the chemical formula \tnote{a}. Default 1\\
    15 &  $Z$ & & Atomic number (number of electron) \\
    16 & $T_0$ & K & Reference temperature for the thermal pressure (see Eq.~\ref{eq:4}). Default 300 \\
    17 & $\alpha_0$ & $10^{-6}$ K$^{-1}$ & The zeroth order coefficient of thermal expansion at a reference pressure P0 \\ 
    18 & $\alpha_1$ & $10^{-6}$ K$^{-2}$ & The first order coefficient of thermal expansion at a reference pressure P0 \\ 
    19 & $\xi$ & & Power law index in the coefficient of thermal expansion (Eq.~\ref{eq:alpha})。 Default 0 \\
    20 & $c_{p0}$ & $10^7$ erg~g$^{-1}$~K$^{-1}$ & Specific heat capacity at constant pressure (see Eq.~\ref{eq:cp})\\
    21 & $c_{p1}$ & $10^7$ erg~g$^{-1}$~K$^{-2}$ & Coefficient for specific heat capacity (see Eq.~\ref{eq:cp})\\
    22 & $c_{p2}$ & $10^7$ erg~K~g$^{-1}$ & Coefficient for specific heat capacity (see Eq.~\ref{eq:cp}) \\
    23 &Debye\_approx &  & Positive number for Debye, otherwise Einstein \\
    24 &thermal\_type & & See table~\ref{tab:temp} \\
    \hline
  \end{tabular}
  \begin{tablenotes}
    \item[a] Number of atoms in the volume of $V/N_A$, where $N_A$ is the Avogadro constant.  The $n$ of ideal gas is the number of atoms per molecule for the purpose of adiabatic index.  For example, $n=2$ for collinear molecules e.g. carbon dioxide. Isothermal atmosphere can be achieved by setting $n=0$.
    \end{tablenotes}
  \end{threeparttable}
\end{table*}

Then the EOS object pointer can be constructed using the EOS constructor that have three arguments.  The first one is a string of the phase description, then is the name of dictionary style double array, and the third argument is \emph{length}, the number of parameters provided.  The name of the EOS object pointer should be the same as the one declared in \emph{EOSlist.h}.

The following shows the EOS of post-perovskite phase of MgSiO$_3$ in the code as an example.

Code in \emph{EOSlist.h}
\begin{lstlisting}[language=C++]
  \\Declare pointers to EOS objects
  extern EOS *Si_PPv_Sakai;
\end{lstlisting}

Code in \emph{EOSlist.cpp}
\begin{lstlisting}[language=C++]
  \\Dictionary style double array
  double Si_PPv_Sakai_array[][2]
  = {{0,4}, {1,24.73}, {2,203}, {3,5.35},
    {5,mMg+mSi+3*mO}, {8,848}, {9,1.47},
    {10,2.7}, {11,0.93}, {15,5}};

  \\ Create a EOS object pointer
  EOS *Si_PPv_Sakai
  = new EOS("Si PPv (Sakai)",
  Si_PPv_Sakai_array,
  sizeof(Si_PPv_Sakai_array)/2
  /sizeof(Si_PPv_Sakai_array[0][0]));
\end{lstlisting}

For RTpress EOS framework, a list of additional polynomial fitting terms $b_n$ , in the unit of erg~mol$^{-1}$, are required (see Eq.~\ref{eq:bV}). A separate array and its size \emph{blength} are used to construct a RTpress EOS object. The following shows the EOS of liquid MgSiO$_3$ as an example.

Code in \emph{EOSlist.h}
\begin{lstlisting}[language=C++]
extern EOS *Si_Liquid_Wolf;
\end{lstlisting}

Code in \emph{EOSlist.cpp}
\begin{lstlisting}[language=C++]
double Si_Liquid_Wolf_array[][2] 
= {{0,10}, {1,38.99}, {2,13.2}, {3,8.238}, 
{5,mMg+mSi+3*mO}, {9,0.1899}, {10,0.6}, 
{12,-1.94}, {15,5}, {17,3000}};
double Si_Liquid_Wolf_b[] 
= {4.738E12, 2.97E12, 6.32E12, 
-1.4E13, -2.0E13};

EOS *Si_Liquid_Wolf 
= new EOS("Si liquid (Wolf)", 
Si_Liquid_Wolf_array, Si_Liquid_Wolf_b, 
sizeof(Si_Liquid_Wolf_array)/2
/sizeof(Si_Liquid_Wolf_array[0][0]),
sizeof(Si_Liquid_Wolf_b)
/sizeof(Si_Liquid_Wolf_b[0]));
\end{lstlisting}

\section{Modify a built-in EOS in runtime}
\label{app:modify-eos}

Once an EOS is built-into the code, it is possible to modify its parameter in the run-time without the request of recompile the full code.  This feature is useful to rapidly repeat runs with similar EOS parameters, for example when study the impact of the uncertainty of EOS parameters on planet size.

One can modify one parameter of a EOS using
\begin{lstlisting}[language=C++]
void EOS::modifyEOS
(int index, double value),
\end{lstlisting}
where index is listed in Table~\ref{tab:param}. Or using
\begin{lstlisting}[language=C++]
void EOS::modifyEOS 
(double params[][2], int length)
\end{lstlisting}
to modify multiple parameters at once, where length is the number of parameters that need to be modified and params is the dictionary style 2D double array (see Appendix~\ref{app:eos-struct}).

If the phase boundary between phases are independent of temperature and purely determined by the pressure, e.g. the phase boundary commonly assumed between ice VII and ice X, the phase transition pressure within a layer can also be modified using
\begin{lstlisting}[language=C++]
void set_phase_highP
(int k, double *start_pressure, 
EOS** phase_name),
\end{lstlisting}
where {\it k} is the number of phases involved in this modification, {\it phase\_name} is an array of EOS pointers of these {\it k} phases, and {\it start\_pressure} is the array of phase transition pressures between these {\it k} phases in the unit of GPa with a length of {\it k}-1.  If the name of the first EOS in the {\it phase\_name} EOS list matches the name of one of the original EOS of this layer, the first input EOS will replace the original one. Otherwise, the first EOS in the list will be ignored in the calculation.

External EOS function, entropy function, or temperature gradient function can be modified using
\begin{lstlisting}[language=C++]
void modify_extern_density
(double (*f)(double P, double T)),
void modify_extern_entropy
(double (*g)(double rho, double T)),
\end{lstlisting}
 or
\begin{lstlisting}[language=C++]
void modify_dTdP
(double (*h)(double P, double T))
\end{lstlisting}
respectively.

\end{CJK*}
\end{document}